# Polymer Concentration Regimes from Fractional Microrheology

**Amirreza Panahi [1], Di Pu [1], Giovanniantonio Natale [1, a), and Anne M. Benneker [1, a)**

**[1] Department of Chemical and Petroleum Engineering, 2500 University Drive NW, Calgary, Alberta, T2N 1N4, Canada.**

## Abstract

In this work, a framework for deriving theoretical equations for mean squared displacement (MSD) and fractional Fokker-Planck (FFP) is developed for any arbitrary rheological model. The obtained general results are then specified for different fractional rheological models. To test the novel equations extracted from our framework and bridge the gap between microrheology and fractional rheological models, microrheology of polystyrene (PS) in tetrahydrofuran (THF) solutions at several polymer concentrations is measured. By comparing the experimental and theoretical MSDs, we find the fractional rheological parameters and demonstrate for the first time that the polymer concentration regimes can be distinguished using the fractional exponent and relaxation time data because of the existence of a distinct behavior in each regime. We suggest simple approximations for the critical overlap concentration and the shear viscosity of viscoelastic liquid-like solutions. This work provides a more sensitive approach for distinguishing different polymer concentration regimes and measuring the critical overlap concentration and shear viscosity of polymeric solutions, which is useful when conventional rheological characterization methods are unreliable due to the volatility and low viscosity of the samples.

## I. Introduction

Characterizing the rheological behavior of soft materials is crucial in various industries including manufacturing [1] and food [2]. Conventional rheometers have limitations in low-viscosity samples at high frequencies, volatile samples, and biological samples where the volume of the sample is limited to a few microliters [3]. For such systems, microrheological methods become relevant. Using only a small volume of the sample and microscopy, the rheological behavior of the sample can be obtained up to very large frequencies. Distinguishing polymer concentration regimes is of importance in industrial applications such as electrospinning and 3-D printing [4] because of changes in viscosity, diffusivity, and longest relaxation time in each regime which will affect the production result [5, 6]. Different polymer concentration regimes are classified as dilute, semi-dilute unentangled, semi-dilute entangled, and concentrated [7]. The existence of different regimes originates from the effect of the polymer concentration on the intermolecular interactions and is unique for each system with a specific solvent quality and polymer molecular weight [6]. The dilute and semi-dilute unentangled regimes are separated at the critical overlap concentration where the polymers start to interact hydrodynamically with each other [8]. The semi-dilute unentangled and semi-dilute entangled regimes are separated at the entanglement onset concentration where the entanglements become elastically effective [9]. Various models have been used previously to describe the polymer dynamics in these concentration regimes such as Rouse, Zimm, and reptation models, where scaling relations are proposed for several parameters including the self-diffusion coefficient and relaxation time [10].

In microrheological methods, the motion of particles is monitored to probe the rheological properties of a system [11, 12]. Passive microrheology is performed by monitoring the thermal motion of tracer particles [13], where generally the multiple-particle tracking (MPT) method is used to capture the motion of tracer particles [14]. The frequency range that can be accessed from the MPT method depends on the capture frame rate. Two methods of passive microrheology are used for viscoelastic characterization: one-point microrheology (1P-microrheology) and two-point microrheology (2P-microrheology) [11]. 1P-microrheology is based on calculating the mean-squared displacement (MSD) of individual particles while 2P-microrheology acquires the MSD from the cross-correlation of particles [15]. 1P-microrheology is less favored because the presence of the particles can change the local structure of the medium, leading to the underestimation of the viscoelastic moduli [16]. 2P-microrheology overcomes this issue by averaging the cross-correlation of particles over multiple lengths [11].

The simplest viscoelastic models are linear viscoelastic models such as Maxwell, Kelvin-Voigt, and linear solid models which are constructed by a combination of Newtonian dashpots and Hookean springs [17]. These models are applicable to data obtained from experimental MSDs when normal diffusion occurs in the solutions. Normal diffusion is Gaussian both in space and time, and the MSD grows linearly in time: $\langle x(t)^2 \rangle = 2dDt$ where $D$ is the Brownian diffusion coefficient and $d$ is the dimensionality of the problem [18]. Anomalous diffusion, a more complex behavior in which the MSD does not scale linearly with time but has a power-law dependence $\langle x \rangle^2 \sim t^\alpha$ [19, 20], exists in many environments. If $0 < \alpha < 1$, the motion is subdiffusive, and if $1 < \alpha < 2$, the motion is superdiffusive [21]. Superdiffusion is generally a result of active motion in the sample while subdiffusion is ubiquitous in condensed-phased systems especially when crowding effects play a role in the particle motion [22]. Subdiffusion can happen when either non-locality in space or time is present. The non-locality in space is related to a class of Markovian stochastic processes called stable Levy motion where arbitrarily long jumps are allowed (Levy flights) [23]. In the current work, we specifically focus on non-locality in time and use the fractional Brownian motion (FBM) model [24] to capture anomalous dynamics of particle motion in a polymeric solution. The FBM has two important properties. First, it is self-similar meaning that any time segment from an FBM trajectory has a behavior similar to any other segments after proper renormalization is performed. Second, it is stationary meaning that its distribution depends on the time lag rather than the starting time [25].

Several mechanical tests have revealed power-law behavior in the creep and relaxation of materials including polymers [17]. For the mathematical description of these materials, fractional rheological models are needed to connect their rheology to the anomalous behavior of their MSDs. In fractional rheological models, stress is related to the fractional derivative of strain [17, 26-28]. Gemant's work was the first instance of this fractional dependence which used a fractional exponent of 0.5 [29]. G.W. Scott Blair pioneered the use of fractional derivatives to describe the creep and stress relaxation of viscoelastic materials [30]. The building blocks of the fractional rheological models are called springpots or Scott-Blair (SB) elements, demonstrating a behavior between Hookean springs and Newtonian dashpots [26, 31]. Springpots can be combined in series or in parallel to obtain the fractional counterparts of the linear viscoelastic models, including the fractional Maxwell model (FMM) and the fractional Kelvin-Voigt model (FKM) [17, 26-28]. Choosing a specific rheological model will affect the two theoretical descriptions that can be used to tackle microrheology. First, the viscoelastic drag term in the equation of motion is affected by the relaxation function derived from the rheological constitutive equation (RCE). Second, the constitutive equation of probability density, which can be derived directly from the equation of motion, is also affected by the relaxation function. Probability density or the probability distribution of particle displacements, denoted as $p(x, t)$, describes the probability of finding a particle at a specific displacement at a specific time. Previously, Santamaria-Holek [32] utilized a general sub-diffusive memory kernel to describe microrheology results with the two aforementioned theoretical descriptions. Fractional rheology in conjunction with microrheology was used previously to capture the material states of cytoplasm during mitosis [33], subdiffusive behavior of optically-trapped cancerous and non-cancerous human cells [34], and a range of solutions from glycerol mixtures to silica gels [35].

Here, we provide a generalized theoretical equation for the MSD for a generic relaxation modulus. We then specify the generalized results to extract new equations for the FMM and a specific case of the FMM. Then, experimental MSDs, obtained from 2P-microrheology, are fitted with the theoretically-derived equation for solutions of various polymer concentrations to extract relevant fractional rheological data. Finally, we suggest novel and more sensitive approaches for differentiating polymer concentration regimes, estimating shear viscosity in liquid-like solutions, and estimating the critical overlap concentration through a combination of theory and microrheology for solutions that are difficult to characterize through conventional methods.

## II. Theory

### A. General MSD Derivation from the Langevin Equation

The Langevin equation is written as a force balance on a particle undergoing Brownian motion in a fluid medium. In this subsection, we derive a general MSD from the Langevin equation which can be utilized for any desired viscoelastic model. For a particle immersed in a viscoelastic fluid, the one-dimensional generalized Langevin equation (GLE) is [36, 37]:

$$m\frac{d\dot{x}}{dt} = -\int_0^t \zeta(t-u)\dot{x}(u)du + F_R(t) + F_{ext} \tag{1}$$

where $m$ is the mass of the particle, $\zeta$ is the friction kernel, and $F_{ext}$ is an externally applied constant force. The considered external force in this work can be a result of a constant electric or gravity field and is independent from the dynamics of the motion of the particle. The functional form of other external forces which depend on spatial particle position and time should be known before the solution as they directly affect the GLE correlators. $F_R$ is the random force which has a zero mean ($\langle F_R \rangle = 0$), and follows the fluctuation-dissipation theorem (FDT) [38, 39].

$$\langle F_R(t)F_R(t')\rangle = C(|t-t'|) = k_BT\zeta(t-t') \tag{2}$$

where $C(t)$ is the correlation function of noise, $k_B$ is the Boltzmann constant, and $T$ is absolute temperature. In the case of a Markovian process without memory effects, the friction kernel is a Dirac delta function [32, 40]. The friction kernel is related to the relaxation modulus according to $\zeta(t-t') = 6\pi RG(t-t')$, where $G$ is the relaxation modulus connecting the stress and strain tensors, and $R$ is the particle radius. Eq. (1) can be solved by Laplace transforming to obtain $x(t)$. Assuming $x(0) = 0$ and using the zero mean rule for the internal noise and the equipartition theorem in one dimension [41], we obtained the general MSD as:

$$\langle x^2 \rangle = 2k_B T I(t) + F_{ext} I(t)[2m\dot{x}(0)H(t) + F_{ext} I(t)] \tag{3}$$

where the GLE correlators $h(t)$, $H(t)$, and $I(t)$ are defined as follows:

$$h(t) = L^{-1}\left\{\frac{1}{ms + 6\pi R\tilde{G}(s)}\right\} \tag{4.a}$$

$$H(t) = \int_0^t h(\tau)d\tau = L^{-1}\left\{\frac{1}{ms^2 + 6\pi R\tilde{G}(s)s}\right\} \tag{4.b}$$

$$I(t) = \int_0^t H(\tau)d\tau = L^{-1}\left\{\frac{1}{ms^3 + 6\pi R\tilde{G}(s)s^2}\right\} \tag{4.c}$$

The MSD equation can also be derived from the constitutive equation of probability distribution for a chosen relaxation modulus associated with a specific rheological model, which is discussed in the next subsection.

**B. The Equivalent Fokker-Planck Representation of the GLE**

The Fokker-Planck (FP) equation is the constitutive equation for the probability distribution of particle displacements $p(x, t)$. The FP equation corresponding to the anomalous diffusion is called the fractional Fokker-Planck (FFP) equation, and it has been derived in the literature for different anomalous transport models including the continuous time random walk (CTRW) model [42-45]. The endeavors in the literature where the FP equation associated with the GLE is derived either use the characteristic function method [46] or the Kramers equation associated with the joint probability distribution [47]. Here, we used the characteristic function method following the work of Wang [46]. We derived the general FP equation corresponding to Eq. (1) as follows (see details in Appendix A):

$$\frac{\partial p(x,t)}{\partial t} = \frac{\partial}{\partial x}\{[-m\dot{x}(0)h(t) - F_{ext}H(t)]P(x,t)\} + \frac{\partial^2}{\partial x^2}\{[k_B TH(t)(1 - mh(t))]P(x,t)\} \tag{5}$$

Eq. (5) demonstrates that several complex forms of the FP equation associated with different viscoelastic models can be derived accordingly. Since the form of $h(t)$ and $H(t)$ are unknown at this point, we cannot simplify the above equation. Once a viscoelastic model is chosen and the associated FP equation is known, the MSD equation can be derived from the Laplace-Fourier transformed probability distribution according to:

$$\langle x \rangle^2(s) = -\frac{\partial^2 \tilde{p}(\omega,s)}{\partial \omega^2}\Big|_{\omega=0} \tag{6}$$

**C. Deriving the MSD and the FFP for the Fractional Maxwell Model**

Due to the presence of subdiffusive behavior in our experimental data, we focus on fractional rheological models and examine how the relaxation modulus affects the MSD. The building units of fractional rheological models are springpots, which are classified with $\alpha$ and $\mu_\alpha$. $\alpha$ is the fractional exponent while $\mu_\alpha$ is a physical parameter with the units of $Pa.s^\alpha$ which can be interpreted as a measure of the firmness of the material. When $\alpha = 0$, $\mu_\alpha$ is equal to the spring constant while $\alpha = 1$ restores a Newtonian viscosity. In the literature, $\mu_\alpha$ has been hypothesized as an apparent viscosity and is related to the Newtonian viscosity by $\mu_\alpha = \eta_\alpha \tau^{\alpha-1}$ [27, 28, 35] where $\tau$ is the relaxation time, and $\eta_\alpha$ is the viscosity of a Newtonian dashpot. Similar to linear viscoelastic models, springpots can be combined in series or in parallel with other springpots, Hookean springs, and Newtonian dashpots to model more complex behaviors of non-Newtonian materials. For instance, the FMM and FKM are obtained when two springpots are put in series and in parallel, respectively [26, 28]. We specify the generalized equations provided in the previous subsections for the general form of the FMM consisting of two springpots in series characterized by $(\alpha, \mu_\alpha)$ and $(\beta, \mu_\beta)$. For this model, without the loss of generality, we assume $\alpha \geq \beta$. The relaxation modulus associated with the FMM is written as [27, 28, 31]:

$$G(t) = \mu_\alpha \tau^{\beta-\alpha} t^{-\beta} E_{\alpha-\beta,1-\beta}\left(-\left(t/\tau\right)^{\alpha-\beta}\right) \tag{7}$$

where $\tau = \left(\mu_\beta/\mu_\alpha\right)^{1/\beta-\alpha}$ is the defined relaxation time for this model and $E_{a,b}$ is the two-parameter Mittag-Leffler function. The one-parameter and two-parameter Mittag-Leffler functions are useful functions in fractional mathematics with the following definitions [26]:

$$E_a(z) = E_{a,1}(z) = \sum_{n=0}^{\infty} \frac{z^n}{\Gamma(an+1)} \quad a > 0 \tag{8.a}$$

$$E_{a,b}(z) = \sum_{n=0}^{\infty} \frac{z^n}{\Gamma(an+b)} \quad a, b > 0 \tag{8.b}$$

where $\Gamma$ is the gamma function. The long-time behavior of the Mittag-Leffler function is obtained according to $E_{a,b}(-z) \underset{z\to\infty}{=} z^{-1}/\Gamma(b-a)$ [38]. Using the relaxation modulus of the FMM, we can derive the associated MSD and FFP equations by taking the Laplace transform of Eq. (7) and substituting the result in Eqs. (4.a)-(4-c). For the polymeric solutions used in our experiments, the relaxation time is ~0.02 s for the highest polymer concentration. Thus, the term $\tau^{\alpha-\beta}s^{2-\beta}$ in the denominator of the obtained GLE correlators can be neglected when compared with the term $s^{2-\alpha}$, which allows us to derive an explicit analytical equation for the MSD. Note that the neglected term can become considerable for solutions with high elasticity such as concentrated polymer solutions and polymer melts at short times, which forces the utilization of numerical inverse Laplace transform. With this simplification and the fractional Laplace transform identities provided by Kimeu [48], it can be written:

$$h(t) = \frac{1}{m}\left[\left(\tau/t\right)^{\alpha-\beta}E_{2-\alpha,1-\alpha+\beta}\left(-\varrho/m\, t^{2-\alpha}\right) + E_{2-\alpha,1}\left(-\varrho/m\, t^{2-\alpha}\right)\right] \tag{9.a}$$

$$H(t) = \frac{t}{m}\left[\left(\tau/t\right)^{\alpha-\beta}E_{2-\alpha,2-\alpha+\beta}\left(-\varrho/m\, t^{2-\alpha}\right) + E_{2-\alpha,2}\left(-\varrho/m\, t^{2-\alpha}\right)\right] \tag{9.b}$$

$$I(t) = \frac{t^2}{m}\left[\left(\tau/t\right)^{\alpha-\beta}E_{2-\alpha,3-\alpha+\beta}\left(-\varrho/m\, t^{2-\alpha}\right) + E_{2-\alpha,3}\left(-\varrho/m\, t^{2-\alpha}\right)\right] \tag{9.c}$$

$\varrho = 6\pi R\mu_\alpha$ is a constant and will be kept throughout the derivation. The first term in the above equations shows the dependence of the GLE correlators to the relaxation time of the medium and is a direct result of our work. In this work, we performed passive microrheology experiments, which makes the second term on the right-hand side of Eq. (3) zero. By substituting Eq. (9.c) into Eq. (3), a new equation for the MSD is readily derived:

$$\langle x^2 \rangle = \frac{2k_BT t^2}{m}\left[\left(\tau/t\right)^{\alpha-\beta}E_{2-\alpha,3-\alpha+\beta}\left(-\varrho/m\, t^{2-\alpha}\right) + E_{2-\alpha,3}\left(-\varrho/m\, t^{2-\alpha}\right)\right] \tag{10}$$

The ratio $\varrho/m$ is large for colloidal particles that are diffusing in the overdamped regime, which allows the use of the asymptotic behavior of the Mittag-Leffler function at large arguments. This simplifies the above equation to the following form:

$$\langle x^2 \rangle = 2D_\alpha\left[\frac{\tau^{\alpha-\beta}t^\beta}{\Gamma(\beta+1)} + \frac{t^\alpha}{\Gamma(\alpha+1)}\right] \tag{11}$$

$D_\alpha = k_BT/\varrho$ is the fractional diffusion coefficient and has the unit of $m^2 s^{-\alpha}$. The obtained equation for the MSD in one dimension is applicable to higher dimensions by multiplication to the number of dimensions.

By substituting Eqs. (9.a) and (9.b) into Eq. (5), a new form of the FFP equation associated with the FMM can be obtained. Using the simplification of the Mittag-Leffler function at large arguments and noting that $m/\varrho = m/6\pi R\mu_\alpha = \tau^{1-\alpha}(m/6\pi R\eta_\alpha) = \tau^{1-\alpha}\tau_c$ where $\tau_c$ is the timescale that separates the overdamped and underdamped regimes, the terms that are multiplied by $\tau_c$ become negligible in the overdamped limit. Therefore, it can be written:

$$\frac{\partial p(x,t)}{\partial t} = m(t)L_{VK}[p(x,t)] \quad (12)$$

where $m(t) = \tau^{\alpha-\beta}t^{\beta-1}/\Gamma(\beta) + t^{\alpha-1}/\Gamma(\alpha)$ is the modified singular memory kernel obtained in our work, $L_{VK} = \partial^2[D_\alpha p(x,t)]/\partial x^2 - \partial[F_{ext}\zeta_\alpha p(x,t)]/\partial x$ is the van Kampen operator [49, 50], and $\zeta_\alpha = 1/\varrho$ is the fractional mobility. When the FFP equation is solved, the probability distribution is generally a fractional function such as the Fox H function [51, 52] which has an equivalent series representation [53]; therefore, $L_{VK}[p(x,t)]$ can be represented as a polynomial in time. It is well known that the multiplication of two polynomials can be written as a convolution, which is essential for solving the FFP equation when utilizing the Laplace transform. This leads to the final form of the FFP equation:

$$\frac{\partial p(x,t)}{\partial t} = (\tau^{\alpha-\beta}{}_0^{RL}D_t{}^{1-\beta} + {}_0^{RL}D_t{}^{1-\alpha})L_{VK}[p(x,t)] = \frac{\tau^{\alpha-\beta}}{\Gamma(\beta)}\frac{\partial}{\partial t}\int_0^t \frac{L_{VK}[p(x,t')]}{(t-t')^{1-\beta}}dt' + \frac{1}{\Gamma(\alpha)}\frac{\partial}{\partial t}\int_0^t \frac{L_{VK}[p(x,t')]}{(t-t')^{1-\alpha}}dt'$$

$$(13)$$

where ${}_0^{RL}D_t$ is the fractional derivative in the Riemann-Liouville sense. Eq. (13) is a non-Markovian FP equation where the memory kernels in the integrals are subdiffusive slow-decaying memory kernels. This derivation provides a novel and more complex form of the FFP equation and demonstrates that various forms of the FFP equation associated with other viscoelastic models can be derived using our methodology. As mentioned earlier, the MSD can also be derived from the FFP equation, and we show in Appendix B that solving Eq. (13) would result in the same MSD equation acquired in Eq. (11) from the GLE. The choice between the GLE and FFP methods depends on the problem. In the case of medium heterogeneities, the FFP equation is more convenient and enable approaches such as the diffusing diffusivity method [54, 55].

**D. Extending the General FMM Results to Simpler Fractional Rheological Models**

The general results obtained previously can be extended to simpler variants of the FMM. The first variant is the single springpot or Scott-Blair (SB) model. This is the simplest fractional rheological model and can be directly inferred from the general FMM results by setting $\beta$ and $\mu_\beta$ equal to zero. This would lead to a zero relaxation time for this model. Accordingly, we can simplify the obtained result in Eq. (10) to the fractional MSD equation previously reported in the literature [56] which has an asymptotic behavior equal to the general form of MSD for the subdiffusive transport of a particle $\langle x^2 \rangle = 2D_\alpha t^\alpha/\Gamma(\alpha+1)$ [57-59]. Additionally, the FFP equation associated with this model can be attained by simplifying Eq. (13), which leads to the commonly used form of the FFP equation in the literature [41-44, 58]. We argue here that the previously reported forms for the fractional MSD and FFP equations cannot be used for cases with high elasticity such as gels and are only applicable to cases where the relaxation time of the medium is negligible with respect to the diffusive time scale of the probe.

Next, from the general FMM results we can obtain two special cases known as fractional Maxwell gel (FMG) and fractional Maxwell liquid (FML) models by setting $\beta = 0$ and $\alpha = 1$, respectively. The FMG model is comprised of a springpot and a Hookean spring and predicts a gel-like behavior where both viscoelastic moduli scale according to $\omega^\alpha$. The FMG model converges to the SB model in the long-time limit or at small frequencies. Because of the liquid-like behavior of our samples, we focus on the FML model comprised of a springpot and a Newtonian dashpot and simplify the general equations of the previous subsection. The corresponding elastic and viscous moduli for the FML model are written from the literature [60]:

$$G'(\omega) = g\frac{(\omega\tau)^{2-\beta}cos\left(\pi\beta/_2\right)}{1+(\omega\tau)^{2(1-\beta)}+2(\omega\tau)^{1-\beta}cos\left(\pi(1-\beta)/_2\right)} \qquad (14.a)$$

$$G''(\omega) = g\frac{(\omega\tau)+(\omega\tau)^{2-\beta}sin\left(\pi\beta/_2\right)}{1+(\omega\tau)^{2(1-\beta)}+2(\omega\tau)^{1-\beta}cos\left(\pi(1-\beta)/_2\right)} \qquad (14.b)$$

where $g = \eta_\alpha/\tau$ [61]. The viscoelastic moduli in Eq. (14) will be compared later with the moduli obtained by performing microrheology in our polymeric solutions. By setting $\alpha = 1$, Eq. (10) can be simplified to the following equation:

$$\langle x^2\rangle = \frac{2k_BT t^2}{m}\left[\left(\tau/_t\right)^{1-\beta}E_{1,2+\beta}\left(-\varrho/_m t\right) + E_{1,3}\left(-\varrho/_m t\right)\right] \qquad (15)$$

Eq. (15) has the following asymptotic behavior in the overdamped limit:

$$\langle x^2\rangle = 2D_\alpha\left[\frac{\tau^{1-\beta}t^\beta}{\Gamma(\beta+1)} + t\right] \qquad (16)$$

In the above equation, the springpot constant in $D_\alpha$ reduces to $\eta_\alpha$. Eq. (16) will be fitted with experimental MSDs obtained from particle tracking and predicts a more complex correlation between MSD and fractional exponent than the generic MSD equation for the subdiffusive particle motion. We also note that setting $\beta = 0$ in Eq. (16) leads to the MSD equation for the classical Maxwell model comprised of a Hookean spring and a Newtonian dashpot combined in series. Finally, the FFP equation associated with the FML model can be obtained by setting $\alpha = 1$ in Eq. (13).

### E. 2P-microrheology

In 2P-microrheology, the motion of two particles is correlated along their centerlines. Two types of correlated motion are investigated in 2P-microrheology; the correlated motion in the direction of the line joining the centers of the particles ($D_{rr}(R_s, \tau_l)$) and the correlated motion perpendicular to this line ($D_{\theta\theta}(R_s, \tau_l)$) where $\tau_l$ is the lag time and $R_s$ is the separation distance [62]. $D_{rr}(R_s, \tau_l)$ is related to the 2P-microrheology MSD through rescaling using a geometric factor [11]:

$$\langle\Delta r^2(\tau_l)\rangle = \frac{2R_s}{R}D_{rr}(R_s, \tau_l) \qquad (17)$$

This equation pinpoints the MSD that is extracted from the particle tracking experimental data and will be fitted with the theoretical MSD equation of the FML model (Eq. 16). In an isotropic, incompressible continuum medium, the viscoelastic spectrum is calculated from the unilateral Laplace transform of $\langle\Delta r^2(t)\rangle$ using the generalized Stokes-Einstein relation (GSER) [12]:

$$\tilde{G}(s) = \frac{k_BT}{\pi Rs\langle\Delta\tilde{r}^2(s)\rangle} \qquad (18)$$

Literature has discussed that for nanoparticles diffusing in a polymer solution, where the particle size is smaller or in the same order of magnitude as the solution's correlation length, strong deviations from the GSER formula are observed, and particles diffuse according to the local viscosity at the microscopic scale [7, 63]. On the other hand, it has been shown experimentally that when the particle size is much larger than the correlation length of the solution, the GSER equation is recovered and the particles experience macroscopic viscosity of the solution [64, 65]. Since in our work the tracer particles are much larger than the solutions' correlation length, the GSER is assumed to be valid throughout the microrheology calculations.

## III. Material and Methods

The experimental system for performing the microrheological characterization consists of a square optically transparent borosilicate glass capillary tube with an inner channel width of 200 $\mu$m and a wall thickness of 100 $\mu$m (Product 8320, VitroCom). Sample is loaded into the capillary through capillary motion. After the capillary is filled with the sample, the two ends of the capillary tube are closed using a UV-curing sealant (UV25, Master Bond Inc. cured at 365 nm wavelength for ~10 seconds) to prevent undesired evaporation which could lead to convective flow in the capillary during experiments. The sample is examined under an inverted microscope (Eclipse Ti2-A, Nikon) equipped with a high-speed camera (iDS uEye, iDS) and an X40/0.60 objective (S Plan Fluor ELWD, Nikon). A Nikon LWD 0.52 condenser lens and an LED light source are used to illuminate the sample. The motion of particles is recorded in the form of an image sequence with an acquisition rate of 100 fps for 30 seconds for each test. Experiments are conducted at room temperature (23 ± 1 °C) for twelve different polymer concentrations ranging from dilute to semi-dilute unentangled regimes. To test repeatability, all tests are done at least three times, and in each recording, approximately ten particles are tracked. To minimize the influence of dynamic errors, the ratio between the exposure time and the lag time in the experiments is kept at around 0.1 [14]. The recorded image sequences are post-processed in the open-source software Fiji ImageJ [66] where background noise is subtracted and particle tracks are determined using the Trackmate plugin [67]. The ensemble and time-averaged MSDs are calculated using codes developed in MATLAB (R2021b, Mathworks). During post-processing of the acquired data, the possible global drift in the data is removed for each case by subtracting the average vector displacement of the particles between the consecutive images from each particle's displacement [62]. The static noise in the system is eliminated by tracking a stationary particle on a glass slide. The obtained MSD is then subtracted from the experimental MSDs to eliminate the effect of localization errors [14]. For weakly subdiffusive cases, experimental MSDs provide the most accurate data when long trajectories with small maximum time lags (around 10 to 20 times the time resolution) are chosen [68].

For polymeric solutions, polystyrene PS615 (Americas Styrenics LLC.) is dissolved in Tetrahydrofuran (THF, Millipore Sigma). Additionally, monodisperse $SiO_2$ microspheres (Glantreo Ltd.) with a nominal diameter of 1 $\mu$m and hydrophobic surface functionality are used. The inherently hydrophilic silica particles are made hydrophobic for stability in the hydrophobic medium by heating the particles at 600 °C to remove the physically adsorbed water and hydroxyl groups from the hydroxylated surface of silica, leaving only siloxane bridges [69]. Monodispersity of the particles, which is crucial for microrheology results through the geometric factor in Eq. (17), is confirmed through images acquired from scanning electron microscopy (SEM, Phenom ProX, Nanoscience Instruments). From a sample of more than 140 particles, the particle diameter was determined to be 0.98 ± 0.07 $\mu$m, indicating that the average diameter is close to the nominal diameter, and the particles have a relatively low polydispersity ($CV = 7.12\%$)

For the THF/PS solution, the critical overlap concentration can be obtained from the molecular weight of the polymer according to $c^*_{THF} = 1/[\eta]_{THF} = 1/0.011M^{0.725}$ [8] where $[\eta]_{THF}$ is the intrinsic viscosity of PS in THF in mL/g, and 0.011 and 0.725 are the associated Mark-Houwink constants. The Mark-Houwink exponent is related to the Flory exponent according to $a_{MH} = 3\vartheta - 1$ [10]. This correlation is useful for the evaluation of the Flory exponent in our solutions, yielding $\vartheta$ =0.575 which is close to the expected value of the Flory exponent in good solvents at $\vartheta$ =0.588. The molecular weight of the polymer molecules was measured using the diffusion-ordered NMR spectroscopy (DOSY) method. In this method, the diffusion coefficient of polymer molecules is accurately measured and correlated with molecular weight according to $D_p = b'M^{-\nu}$. For the DOSY measurements, the polymers were dissolved in deuterated chloroform ($CDCl_3$), and the viscosity-corrected constants for this mixture are provided in the work of Voorter et al. [70]. For PS615, the average molecular weight is determined as 263,573 g/mol which yields a critical overlap concentration of $c^* = 1.210$ wt%. The entanglement onset concentration of the solution is obtained by $c_e \approx \rho_p \left( M_e^0 / M \right)^{3\vartheta - 1}$ where $M_e^0 = 17,000$ g/mol is the average molecular weight between entanglements in the undiluted polymer [10, 71]. As a result, the entanglement onset concentration of the PS615 solution is determined to be $c_e = 10.531$ wt%. The mesh size or correlation length of our semi-dilute THF/PS solutions depends on polymer concentration according to $\xi_{PS} = 2.7c^{-0.68}$ where $\xi_{PS}$ is the solution mesh size in Å and $c$ is polymer concentration in g/mL [72]. Accordingly, the highest mesh size at the critical overlap concentration is calculated to be 5.91 nm. For measuring the solutions' shear viscosities, an Ubbelohde glass viscometer (9721-K50, CANNON Instrument Company) was used since the high volatility and low viscosity of the solutions prevented conventional viscometers from producing reliable data. The glass viscometers measure the efflux time of a specific volume of the solution. In these viscometers, the driving force of the flow is the hydrostatic pressure of the liquid column inside the viscometer [73]. The constant of the viscometer, which depends on the viscometer's geometrical characteristics, was provided by the manufacturing company and is used together with the efflux time to accurately determine the kinematic viscosity of each solution. The viscosity of every solution was measured three times.

## IV. Results and Discussion

The viscosities measured using the Ubbelohde glass viscometer at several solution concentrations are scaled with the pure solvent viscosity and shown in Fig. 1. The shear viscosity of the solutions increases significantly with polymer concentration. The measured viscosity data is fitted with the phenomenological Phillies equation $\eta_r = \eta/\eta_s = \exp(\chi c^\theta)$ which has a stretched exponential form [74]. Here, $\eta_s$ is the pure solvent viscosity measured to be 0.539 mPa·s while $\chi$ and $\theta$ are the scaling exponents. This equation was suggested on an empirical basis, and it has been used for various tracers (hard spheres, globular proteins, linear and branched chains) in different types of polymer matrices in various concentration regimes up to polymer melts [74]. A heuristic proof by Phillies shows that the drag coefficient exerted by the polymers on the probe changes in a stretched exponential form [75], and the viscosity is directly proportional to this force. The stretched exponential form is valid for cases where the hydrodynamic interactions between freely rotating objects are considered. Since the particle size in our experiments is orders of magnitude larger than the mesh size or tube diameter which are around a few nanometers, the hydrodynamic stretched exponential model applies to our data rather than the models considering topological effects such as probe caging. Previously, it was shown that $\theta$ depends on the polymer molecular weight according to $M^{-1/4}$ and changes from one at the low molecular weight limit ($M = 1 \times 10^5$ g/mol) to 0.5 at the high molecular weight limit ($M = 5 \times 10^5$ g/mol) [75]. In this range, the polymers exhibit a transition from small chains to blob-model statistics. The utilized polymer molecules in our solutions are in the intermediate molecular weight range leading to $\theta =0.71$, which is the reported value for the THF/PS mixture with $M = 1.7 \times 10^5$ g/mol [75]. In addition, $\chi$ has a linear dependence on the polymer's molecular weight, and our obtained value is 17.1 for $\theta =0.71$ which is comparable to the reported value in the literature at 16.1 for $\theta =0.75$ [75]. Note that the unit used for polymer concentration directly affects $\chi$, which is dimensionless in our case. Our viscosity data also agrees with the reported concentration scaling laws [76-79] within each polymer concentration regime. These scaling laws and their utilization for our data are discussed in Appendix C.

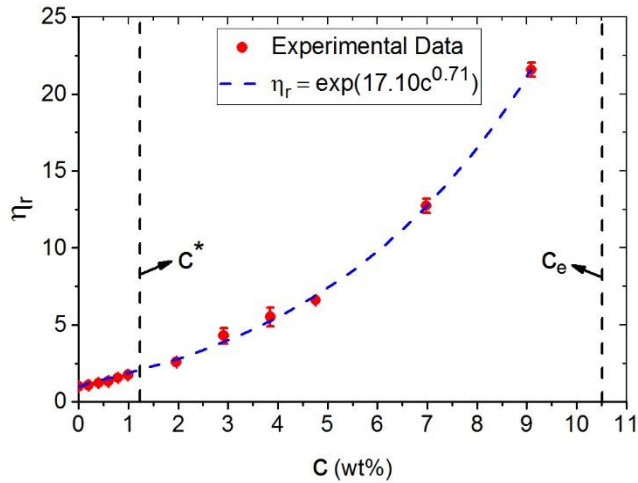

Figure 1. Measured viscosity of the THF/PS solutions as a function of concentration scaled with the solvent viscosity. The dashed line presents the fitted stretched exponential function suggested by Phillies [74]. Critical overlap and entanglement onset concentrations are indicated by vertical lines.

In Fig. 2, experimental MSDs obtained from 2P-microrheology (Eq. (17)) are fitted with the theoretically derived MSD of the FML model (Eq. (16)) for the polymeric solutions in the dilute (Fig. 2.a) and semi-dilute (Fig. 2.b) regimes. It should be noted that the theoretical derivation was done for a one-dimensional case while our experiments are analyzed in two dimensions. Thus, the result of Eq. (16) is multiplied by the dimensionality of the problem. Furthermore, fitting the MSDs with both the original (Eq. 15) and the overdamped (Eq. 16) versions resulted in identical fitting constants even for the highest polymer concentrations; therefore, we have only shown the fittings with the overdamped version (Eq. 16) in the figure to avoid redundancy. The MSDs in Fig. 2 are scaled with the square of the particle radius while the time lags are non-dimensionalized with $\tau_d = R^2/D_s = R^2 \zeta^s/k_B T$. $\tau_d$ is the Brownian diffusive time scale over which the particle diffuses a distance equal to its radius [80]. As mentioned earlier, the GSER equation is valid for our problem, and we have used the Stokes-Einstein (SE) equation to calculate the particle diffusivity in pure solvent $D_s$. From the SE equation it directly follows $\zeta^s = 6\pi R \eta_s$, which is the Stokes dissipative friction force coefficient applied to a particle diffusing in the pure solvent. As can be seen in Fig. 2, the experimental MSDs are well described with the theoretical equation, especially for the highest concentrations at low times. It can be observed that the MSD of particles significantly reduces with increasing polymer concentration in both regimes. This indicates that an increased presence of polymer molecules increases viscosity, and it also causes a crowding effect which decreases the available free volume for particle diffusion, which are in line with expectations. From Eq. (16), it can be inferred that the first term, which is proportional to $\tau^{1-\beta}t^{\beta}$, is dominant at low time lags while the second term, which is proportional to $t$, is dominant at long time lags. The initial subdiffusivity in the data that becomes more noticeable with increasing polymer concentration is due to the effect of increased elasticity and resulting larger relaxation times at higher polymer concentrations, which influences the MSDs through the $\tau^{1-\beta}$ term in Eq. (16). The MSD behavior captured here is in qualitative agreement with the previously discussed coupling between particle dynamics and polymer relaxation modes, which states that the particles that are larger than the entanglement length of the polymer chain show a subdiffusive behavior at short times and a diffusive behavior at long times [7]. This change in the MSD behavior was experimentally shown to occur at time scales comparable to the relaxation time of the polymer [65]. We here provide a single equation to capture the entire MSD of the particles and show theoretically that the initial subdiffusive behavior depends on the relaxation of polymer chains over the particle surface. Our observations show that the previously-utilized subdiffusive fitting in low time lags [65] cannot determine the fractional exponent accurately as this parameter constantly increases before reaching its asymptotic value at long time lags. As mentioned earlier, for the cases where subdiffusive behavior is expected, the optimal time lag for accurate fittings with time-averaged MSDs should be a small fraction of the entire trajectory [68]. For our analysis, a maximum time lag of 0.75s averaged across the entire recorded data was chosen, which produced consistent fittings throughout all sets of data. The fitting constants from these MSDs yield the fractional exponent, relaxation time, and fractional diffusion coefficient. The fractional diffusion coefficient can be used to calculate the viscosity of the Newtonian dashpot.

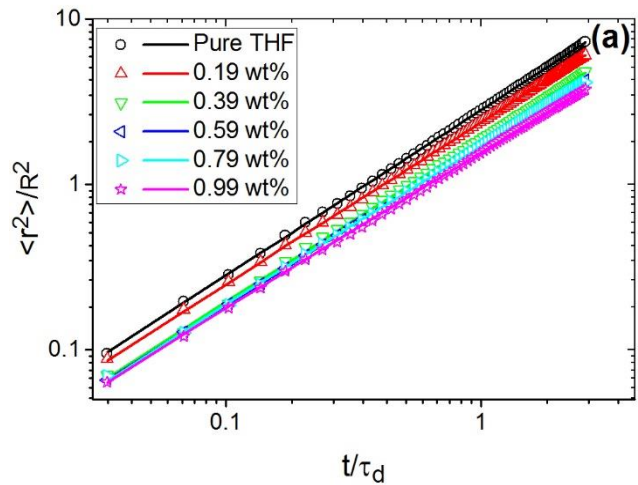

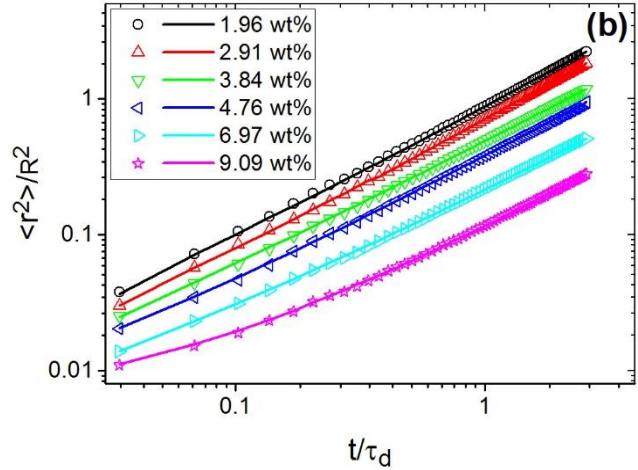

Figure 2. Experimental MSDs obtained from 2P-microrheology fitted with the theoretical MSD equation (Eq. (16)) for solutions in (a) the dilute regime and (b) the semi-dilute regime. The MSDs are scaled with the square of the particle radius while the time lags are non-dimensionalized with the Brownian diffusive time scale ($\tau_d$). The points are the experimental data while the lines are the fitting results.

Next, we examine how well the FML model describes the experimentally determined rheological behavior of the solutions. In Fig. 3, the viscoelastic moduli acquired from experimental MSDs according to Eq. (18) are plotted together with the viscoelastic moduli of the FML model obtained from Eq. (14). To directly compare the theory and experiment in both regimes, one case for each of dilute and semi-dilute regimes is shown in Fig. 3. The plotted viscoelastic moduli are non-dimensionalized by $\eta_s\dot{\gamma}_0$, where $\dot{\gamma}_0$ is the characteristic strain rate. The characteristic strain rate that is chosen must fall within the range of deformation rates that exist in the flow [81]. A characteristic strain rate can be obtained from the reciprocal of the Brownian diffusive time scale $1/\tau_d = D_s/R^2$, which is the deformation rate for flows in the vicinity of diffusive Brownian particles. Using the Brownian diffusive time scale, the angular frequency values are rescaled according to a Deborah number defined as $De = \omega\tau_d$ [82]. It is observable in Fig. 3 that the FML model formed by a springpot and a Newtonian dashpot has captured the rheological behavior of the polymeric solutions accurately over the entire $De$ range with a similar convergence between the viscoelastic moduli at higher $De$ values. Note that the values smaller than a certain percentage of $G(s)$ are eliminated because of their small signal to noise ratio, which has resulted in the elastic moduli data in the lower $De$ range to be clipped. It can be further observed from Fig. 3 that the elastic and viscous moduli have a terminal behavior according to $G' \propto \omega^{2-\beta}$ and $G'' \propto \omega$, while the viscoelastic moduli are generally known to scale according to $G' \propto \omega^2$ and $G'' \propto \omega$ in the terminal region [83]. The observed scaling of viscoelastic moduli in the FML model can also be inferred from Eq. (14) and shows that as the fractional exponent increases with polymer concentration, the elastic modulus deviates more from its known scaling exponent within the terminal relaxation zone. For low-concentration solutions in the dilute regime, relaxation times are small (or the crossover of viscoelastic moduli happens at high $De$ values). By increasing polymer concentration, both moduli increase in magnitude while the crossover point happens at lower $De$ values. This is due to the significant rise in relaxation time when increasing polymer concentration, which will be discussed later (see Fig. 6).

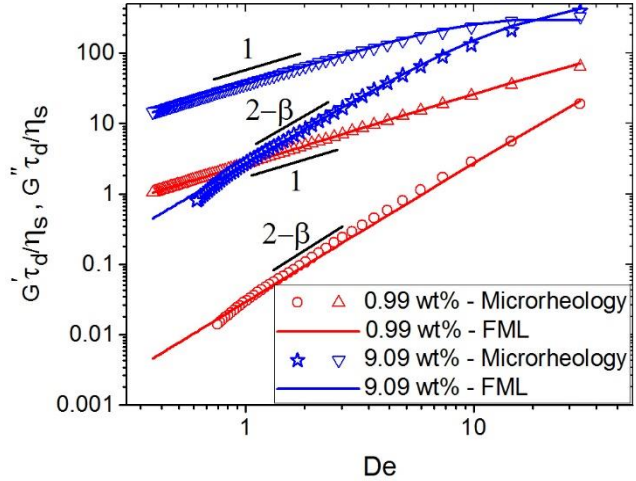

Figure 3. Comparison between the non-dimensionalized viscoelastic moduli obtained directly from the experimental MSDs and the FML model results calculated from Eq. (14) for one concentration in (a) the dilute regime (c = 0.99 wt%, red lines and symbols, ○: elastic modulus, Δ: viscous modulus) and (b) the semi-dilute regime (c = 9.09 wt%, blue lines and symbols, ☆: elastic modulus, ▽: viscous modulus) plotted against the Deborah number. The frequency scalings within the terminal region are also shown on the plots.

Subsequently, the fractional exponents, viscosities of Newtonian dashpots, and relaxation times acquired from MSD fittings are shown against polymer concentration in Figs. 4, 5, and 6, respectively. In Fig. 4, the fractional exponent data in the dilute and semi-dilute regimes have been fitted with two distinct lines, where a significant change in the slope is found between the dilute and semi-dilute regimes. In the dilute regime, the fractional exponent increases more dramatically with concentration than in the semi-dilute regime. A similar distinction in the fractional exponent behavior between dilute and semi-dilute regimes is evident in the data provided by Poling-Skutvik et al. [65] for large particles, even though they report the decrease of fractional exponent $\alpha$ from one which is observable in systems that depict a gel-like behavior and are described by the SB or FMG models. In systems with large particles, the caging effects typically observed in colloidal systems [84] cannot be a reason for the anomalous behavior because the particle size is much larger than the mesh size, which is in the order of a few nanometers. The change of slope in the fractional exponent data happens because added polymers are interacting with the colloids through excluded volume interactions. As the critical overlap concentration is reached, polymer molecules start to interact with each other. As a result, their hydrodynamic interactions with the tracer particles are reduced, causing a slower rise in the fractional exponent. According to the predictions derived from scaling laws, the fractional exponent continues to change with polymer concentration before reaching a plateau value [7], where this plateau value was linked to the degree of coupling between particle dynamics and relaxation modes of the polymer [65]. What is significant here is that by using only the fractional exponents attained from experimental MSDs, the change in the polymer concentration regime is clearly captured. Traditionally, these regimes are distinguished by using intrinsic viscosity, and we show here for the first time that the fractional exponent data can provide an alternative when the intrinsic viscosity of the solution is not available or hard to acquire. The existence of two lines with specific equations in both dilute and semi-dilute regimes can provide an alternative approach for estimating the critical overlap concentration of a polymer solution. The fitted lines shown in Fig. 4 cross at the concentration $c^*_{est}$ =0.992 wt% which predicts the critical overlap concentration with an error of 17.9% when compared to the value calculated from the intrinsic viscosity and its associated Mark-Houwink constants. More measurements close to the critical overlap concentration could further improve the accuracy of this prediction.

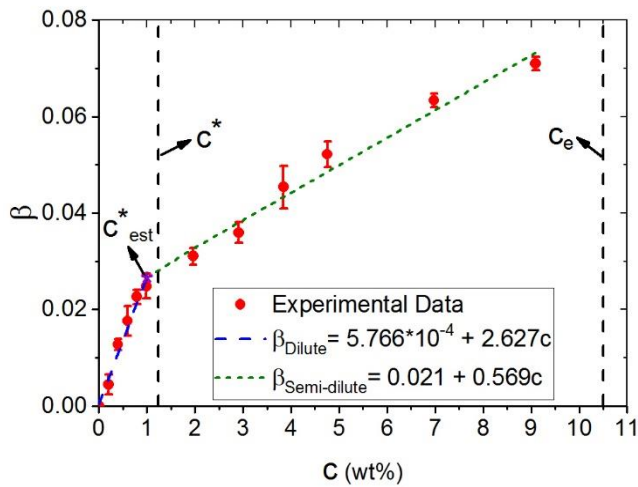

Figure 4. Fractional exponents attained from fitting the experimental MSDs with Eq. (16) for several polymer concentrations. The equations of the fitted lines in each regime are provided in the legend. The point where the fitted lines cross is highlighted, and critical overlap and entanglement onset concentrations are indicated by vertical lines.

In Fig. 5, $\eta_\alpha$ values are scaled with their corresponding value in the pure solvent. $\eta_\alpha$ substantially rises with increasing polymer concentration. The relative changes of $\eta_\alpha$ are captured in our work across all polymer concentration regimes with a stretched exponential function similar to the Phillies equation used for the shear viscosity data. From correlation $g = \eta_\alpha/\tau$ in Eq. (14), it can be inferred that the viscoelastic moduli are directly proportional to the viscosity of the Newtonian dashpot in the FML model. Additionally, the stretched exponential increase in $\eta_\alpha$ with polymer concentration is dominant over the power law increase in relaxation time (see Fig. 6), which causes the incremental increase in the viscous and elastic moduli of the solutions observable in Fig. 3. According to $\mu_\alpha = \eta_\alpha \tau^{\alpha-1}$ for $\alpha$ =1, the value of $\eta_\alpha$ can provide an average frequency-independent estimation of shear viscosity in viscoelastic liquids modeled by the FML. Thus, determination of $\eta_\alpha$ through MSD fittings for such systems is an alternative approach to get the shear viscosity value. Note that the Phillies equation scaling exponents for both viscosity and $\eta_\alpha$ are almost equal. This is because the term $\tau^{\alpha-1}$ that correlates these values is one in the FML model. Therefore, the correlation $\eta/\eta_s = \eta_\alpha/\eta_{\alpha_s}$ provides a reasonable estimation of shear viscosity when the solvent viscosity and $\eta_{\alpha_s}$ data are available. The obtained shear viscosity is at the frequency (shear rate) where the solution viscosity is measured, which depends on the efflux time and the geometric characteristics of the glass viscometer.

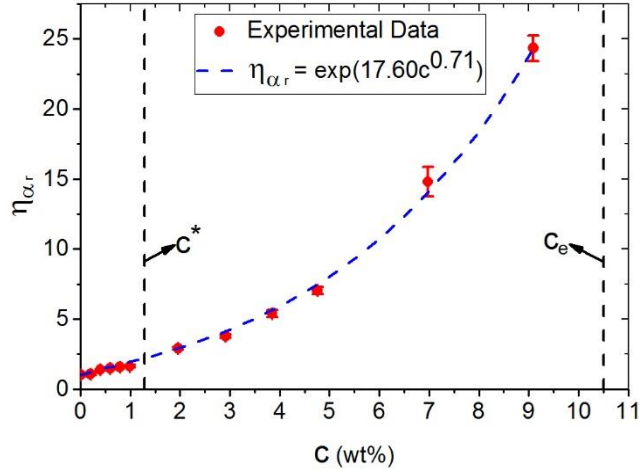

Figure 5. Viscosity of the Newtonian dashpot data obtained from fitting the experimental MSDs with Eq. (16) for several polymer concentrations scaled with the corresponding value in the pure solvent. The dashed line presents the fitted stretched exponential function. Critical overlap and entanglement onset concentrations are indicated by vertical lines.

Finally, the relaxation times of our solutions are non-dimensionalized with the Brownian diffusive time scale and depicted in Fig. 6 against polymer concentration. The different polymer concentration regimes are clearly distinguishable from the scaled relaxation time data. For solutions in the dilute regime, the relaxation time is close to zero because the polymer chains are too dilute to impose any elasticity in the fluid. Once the semi-dilute region is reached, the relaxation times become non-negligible with respect to $\tau_d$, and the data demonstrates a power law increase within the semi-dilute unentangled regime. It is known that the Rouse model can capture the dynamics of polymer solutions in the semi-dilute unentangled regime on length scales larger than the hydrodynamic screening length [10]. As a result, we verified the obtained relaxation times from our work against the longest Rouse relaxation times written as [6, 10]:

$$\tau_R \approx \frac{\eta_s b^3}{k_B T} N^2 c^{2-3\vartheta/3\vartheta-1} \tag{19}$$

where $b$ is the size of a Kuhn segment, and $N$ is the number of Kuhn segments in the polymer chain. The measured persistence length $l_p$ of PS in THF provided by Haidar Ahmad and Striegel [85] is used to estimate the Kuhn length of PS according to $b \approx 2l_p$. The effective Kuhn segment of PS is reported to be around 50 monomers ($\sim$ 5,000 gr/mol) [86], which is used to approximate the number of Kuhn segments in our polymer chain. Referring to the Flory exponent obtained from the Mark-Houwink constant at $\vartheta =0.575$, the Rouse relaxation times are scaled with concentration with a scaling exponent of 0.379. This scaling together with the calculated Rouse relaxation times are shown with a green dashed line in Fig. 6, which shows a reasonable agreement with our experimental data. The observed discrepancy between the curves in Fig. 6 can be explained by the fact that ideal chains are considered in the Rouse model, and complexities such as excluded volume effects, non-local interactions, and chain stiffness are not considered. In the literature, the discrepancy between the relaxation times acquired from experiments and the Rouse theory is addressed by replacing the solvent friction coefficient with an effective friction coefficient when the Markovian dynamics assumption holds [87]. In addition, the Flory exponent can be calculated from the scaling exponent of the power law function fitted with our experimental relaxation time data at 0.465 (Fig. 6 blue dashed line), leading to $\vartheta$ =0.560 which is close to the expected value for this parameter in our solutions. The data follows this power law trend before deviating significantly near the entanglement onset concentration as the formation of entanglements and gradual switching to tube reptation dynamics causes a sharp rise in the time that the polymer molecules require to readjust to the new position of the tracer particle.

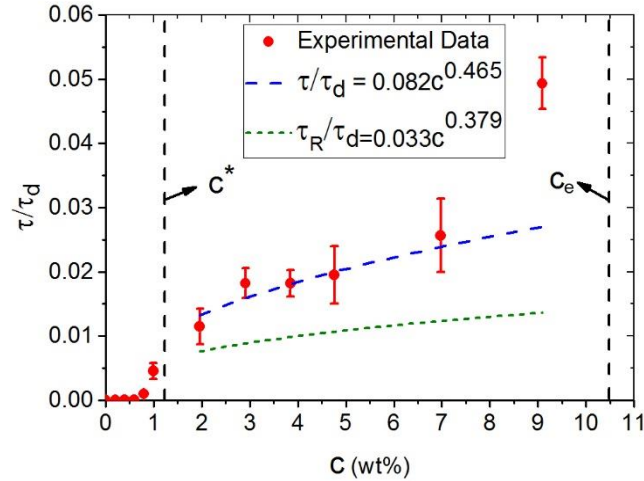

Figure 6. Relaxation time data acquired from fitting the experimental MSDs with Eq. (16) together with the calculated Rouse relaxation times for several polymer concentrations. The relaxation times are non-dimensionalized with the Brownian diffusive time scale ($\tau_d$). The equations of the power law functions in the semi-dilute unentangled regime are provided in the legend. Critical overlap and entanglement onset concentrations are indicated by vertical lines.

## V. Conclusions

In this work, a general theoretical framework for deriving the MSD and FFP equations applicable to any rheological model is developed. The novel equations extracted from this framework are specified for different fractional rheological models and compared to the experimental data obtained from 2P microrheology. Fitting between the MSDs and the theoretical equation provided fractional rheological parameters in several polymer concentrations. We demonstrated for the first time that the fractional exponent data can be used to distinguish polymer concentration regimes and specify the critical overlap concentration due to the existence of two distinct slopes in the dilute and semi-dilute regimes. We also observed that changes in the shear viscosity and the viscosity of a Newtonian dashpot with polymer concentration can be captured throughout the entire tested polymer concentration regimes using stretched exponential functions. Due to the similarity between the scaling exponents of both stretched exponential functions, the correlation $\eta/\eta_s = \eta_\alpha/\eta_{\alpha_s}$ was inferred for viscoelastic liquid-like solutions behaving according to the FML model. Finally, the polymer concentration regimes can be identified through relaxation time data, which exhibits unique behavior in different regimes. It is observed that in the dilute regime, relaxation times are small due to the weak elasticity of the solutions. In the semi-dilute regime, relaxation times become considerable with respect to the Brownian diffusive time scale of the particle and show a power law increase before reaching near the entanglement onset concentration, where a sudden jump in relaxation time is seen. The results of this research provide alternative methods for identifying different polymer concentration regimes and estimating the shear viscosity of polymeric solutions that are difficult to characterize conventionally due to high volatility or low viscosity.

**Conflicts of Interest**

The authors have no conflicts to disclose.

**Acknowledgements**

The authors gratefully acknowledge the financial support from the Canada First Research Excellence Fund (CFREF).

**Appendix A: Derivation of the General FP Equation Using the Characteristic Function Method**

In the probability theory, the probability distribution can be defined according to its characteristic function [46]:

$$\varphi(y,t) = \langle \exp(ixy) \rangle = \int p(x,t)\exp(ixy)\, dx \tag{A.1}$$

where $\varphi(y,t)$ is the characteristic function. By solving Eq. (1), the displacement increment can be written according to the fluctuating force:

$$\Delta x = x(t) - \langle x \rangle = \int_0^t H(t-\tau)F_R(\tau)d\tau = x(t) - x(0) - m\dot{x}(0)H(t) - F_{ext}I(t) \tag{A.2}$$

Since the correlation effects are considered up to the second order for the fluctuating force, the characteristic function is also expressed in terms of the first and second moments. The first moment is equal to the mean, and the higher central moments of order $m$ around the mean are written as $M_m = \langle [x - \langle x \rangle]^m \rangle$ [88]. By expanding $\langle ixy \rangle$ according to the first and second moment ($\langle ixy \rangle = \sum_{m=1}^2 M_m/m!\,(iy)^m$), it can be written:

$$\varphi(y,t) = \exp\left(iy\{x(0) + m\dot{x}(0)H(t) + F_{ext}I(t)\} - \frac{1}{2}\sigma_{xx}y^2\right) \tag{A.3}$$

$\sigma_{xx}$ is defined as [89]:

$$\sigma_{xx} = \langle \int_0^t H(t-t_1)F_R(t_1)dt_1 \int_0^t H(t-t_2)F_R(t_2)dt_2 \rangle \tag{A.4}$$

Referring back to the FDT (Eq. 2) and using its symmetry property ($C(t_1 - t_2) = C(t_2 - t_1)$), the variance $\sigma_{xx}$ can be written as [89, 90]:

$$\sigma_{xx} = 2\int_0^t H(t_1)dt_1 \int_0^{t_1} H(t_2)C(t_1 - t_2)dt_2 = k_B T\left(2I(t) - mH^2(t)\right) \tag{A.5}$$

Next, the time derivative of Eq. (A.3) is taken:

$$\frac{\partial \varphi(y,t)}{\partial t} = \left[iy\{m\dot{x}(0)h(t) + F_{ext}H(t)\} - \frac{1}{2}y^2\frac{d\sigma_{xx}}{dt}\right]\varphi(y,t) \tag{A.6}$$

By taking the time derivative of Eq. (A.5), the relation $d\sigma_{xx}/dt = 2k_B T H(t)\left(1 - mh(t)\right)$ is obtained. Substituting this result in Eq. (A.6) yields:

$$\frac{\partial \varphi(y,t)}{\partial t} = \left[iy\{m\dot{x}(0)h(t) + F_{ext}H(t)\} - y^2 k_B T H(t)\left(1 - mh(t)\right)\right]\varphi(y,t) \tag{A.7}$$

Finally, taking the inverse Fourier transform of the above equation yields the general FP equation reported in Eq. (5).

**Appendix B: Derivation of the MSD from the FFP Equation**

Here, we derive the MSD by solving the obtained integro-differential FFP equation provided in Eq. (13) in the Laplace-Fourier space when the constant external force is zero. The Laplace transform of the Riemann-Liouville fractional derivative is written as [91]:

$$L\{_{0}^{RL}D_{t}^{\alpha}x(t)\} = s^{\alpha}L\{x(t)\} - \sum_{k=0}^{n-1}s^{n-k-1}\frac{d^{(k-1)}}{dt^{(k-1)}}I_{0}^{n-\alpha}x(0^{+}) \tag{B.1}$$

where $n = [\alpha] + 1$, and $I_{0}^{n-\alpha}$ is the Riemann-Liouville fractional integral of order $n - \alpha$. For the initial probability density, if we assume the particle was initially located at the origin, then $p(x,0) = \delta(x)$ [43, 45, 46]. Performing the Laplace transform on Eq. (13) then yields:

$$s\tilde{p}(x,s) - \tilde{p}(x,0) = (\tau^{\alpha-\beta}s^{1-\beta} + s^{1-\alpha})\left\{\frac{\partial^{2}}{\partial x^{2}}[D_{\alpha}\,\tilde{p}(x,s)]\right\} \tag{B.2}$$

where ~ is used to exhibit the Laplace-transformed parameters. By Fourier transforming Eq. (B.2) and noting that $D_{\alpha}$ is constant for a specific solution at a constant temperature, it can be written:

$$\hat{\tilde{p}}(\omega,s) = \frac{1}{s + \{\tau^{\alpha-\beta}s^{1-\beta} + s^{1-\alpha}\}\omega^{2}D_{\alpha}\hat{\tilde{p}}(\omega,s)} \tag{B.3}$$

where ^ is used for the Fourier-transformed parameters. Substituting Eq. (B.3) into Eq. (6) will lead to the following equation:

$$\langle x \rangle^{2}(s) = 2D_{\alpha}\left[\frac{\tau^{\alpha-\beta}}{s^{\beta+1}} + \frac{1}{s^{\alpha+1}}\right] \tag{B.4}$$

Inverse Laplace transforming the above equation will yield the same result obtained in Eq. (11) from the GLE.

**Appendix C: The Shear Viscosity Scaling with Polymer Concentration**

The Huggins equation for the viscosity of homogenous solutions of linear polymers is written as follows [76, 77]:

$$\eta_{sp}(c) = [\eta]c + k_{H}([\eta]c)^{2} + \cdots \tag{C.1}$$

where $\eta_{sp} = (\eta(c) - \eta_{s})/\eta_{s}$ is specific viscosity, $[\eta]$ is intrinsic viscosity, and $k_{H}$ is the Huggins coefficient. For low polymer concentrations, Eq. (C.1) can be approximated up to the first term, yielding a first-order concentration dependence for viscosity in the dilute regime. It has been discussed in the literature that viscosity scaling in semi-dilute unentangled and semi-dilute entangled regimes differ significantly and follow these scaling laws [76-79]:

$$\eta_{sp}(c) \propto \begin{cases} c^{1.25} & semi-dilute\ unentangled \\ c^{15/4} & semi-dilute\ entangled \end{cases} \tag{C.2}$$

In the semi-dilute unentangled regime, scaling exponents in the range of 1.1-1.4 were reported for the solutions of good solvents [76, 77, 79]. As can be seen in Fig. 7, the specific viscosity of our solutions is proportional with polymer concentration within the experimental error bars in the dilute regime. In the semi-dilute unentangled regime, a scaling exponent of 1.606 is obtained which shows that our solutions have a slightly stronger concentration dependence than theoretical predictions. The acquired scaling exponent in the semi-dilute unentangled regime can be related to the Flory exponent according to $\eta_{sp}(c) \propto c^{1/3\vartheta-1}$ [5, 10], leading to $\vartheta = 0.541$ which is in reasonable agreement with the expected Flory exponent in our solutions. The point near the semi-dilute entangled regime slightly deviates from the semi-dilute unentangled scaling line which is expected due to the much stronger concentration dependence for this regime. It is noteworthy that the point where the fitted scaling lines cross provides an estimation of the critical overlap concentration at $c^{*}_{est} = 1.693$ wt%. This estimation has an error of 39.9%, which is significantly higher than the error of our proposed estimation using the fractional exponent values.

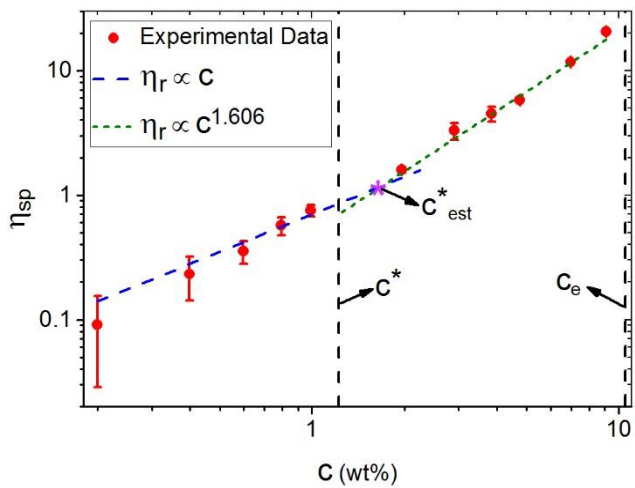

Figure 7. Specific viscosity of the THF/PS solutions as a function of polymer concentration scaled with the viscosity scaling laws. The dashed lines present the fitted scaling laws within each polymer concentration regime. The estimated critical overlap concentration is highlighted at the point where the two dashed lines cross. Critical overlap and entanglement onset concentrations are indicated by vertical lines.